\documentclass[12pt,a4paper]{article}

\usepackage{a4wide}

\newcommand\eqref[1]{(\ref{#1})}
\newcommand\II{I\kern-0.1em I}
\DeclareSymbolFont{AMSb}{U}{msb}{m}{n}
\DeclareSymbolFontAlphabet{\mathbb}{AMSb}
\newcommand\inv{^{\raise.15ex\hbox{${\scriptscriptstyle -}$}\kern-.05em 1}} 
\newcommand\grad[2][]{\mathop{\nabla_{\!#2}^{#1}}\nolimits}	
\newcommand\hatgrad[2][]{\mathop{\widehat\nabla_{\!#2}^{#1}}\nolimits}
\newcommand\st{{}^*\kern-.14em}				
\newcommand\CA{\mathcal{A}}
\newcommand\CB{\mathcal{B}}
\newcommand\CE{\mathcal{E}}
\newcommand\CH{\mathcal{H}}
\newcommand\CL{\mathcal{L}}
\newcommand\CM{\mathcal{M}}
\newcommand\CN{\mathcal{N}}
\newcommand\eps{\epsilon}
\newcommand\set[1]{\mathbb{#1}}				
\newcommand\C{\set{C}}
\newcommand\Z{\set{Z}}
\newcommand\T{\set{T}}
\newcommand\group[1]{\mathop{\kern0pt\mathrm{#1}}\nolimits}	
\newcommand\U{\group{U}}				
\newcommand\SU{\group{SU}}	
\newcommand\SO{\group{SO}}	
\newcommand\SL{\group{SL}}
\newcommand\E{\group{E}}
\newcommand\opname[1]{\mathop{\kern0pt\mathrm{#1}}\nolimits}	
\newcommand\Tr{\opname{Tr}}				
\renewcommand\det{\opname{det}}				
\newcommand\e{\opname{e}}				
\makeatletter						
\renewcommand\section{\@startsection{section}{1}{\z@}%
                                    {-7ex \@plus -1ex \@minus -.2ex}%
                                    {2.5ex \@plus.2ex}%
                                    {\normalfont\large\scshape\centering}}                                   
\renewcommand\subsection{\@startsection{subsection}{2}{\z@}%
                                       {-5ex \@plus -1ex \@minus -.2ex}%
                                       {1.5ex \@plus.2ex}%
                                       {\normalfont\normalsize\scshape}}

\renewcommand\abstract{\section*{\abstractname}}                                   
\newcommand\ack{\section*{\ackname}}                                   
\newcommand\ackname{Acknowledgements}
\renewcommand\appendix{\par
  \setcounter{section}{0}%
  \setcounter{subsection}{0}%
  \renewcommand\thesection{\appendixname~\@Alph\c@section}%
  \renewcommand\thesubsection{\@Alph\c@section.\@arabic\c@subsection}}
\renewcommand\@seccntformat[1]{\csname the#1\endcsname.\quad} 
\long\def\@makecaption#1#2{%
  \vskip\abovecaptionskip
  \sbox\@tempboxa{\textsc{#1}: #2}%
  \ifdim \wd\@tempboxa >\hsize
    \textsc{#1}: #2\par
  \else
    \global \@minipagefalse
    \hb@xt@\hsize{\hfil\box\@tempboxa\hfil}%
  \fi
  \vskip\belowcaptionskip}
\let\keywords\@gobble
\makeatother						
\begin{document}

%
%

\begin{flushright}
\textsc{spin-98/2\\ hep-th/9810116\\ Oktober 1998}
\end{flushright}
\vskip1cm

\begin{center}

{\LARGE\scshape U-Duality of Born-Infeld on the\\ 
   Noncommutative Two-Torus \par}
\vskip25mm

\textsc{Christiaan Hofman} and 
\textsc{Erik Verlinde}\\[3mm]

\textit{Spinoza Institute \textnormal{and} Institute for Theoretical Physics\\
        University of Utrecht, 3584 CE Utrecht}\\[2mm]

\texttt{hofman@phys.uu.nl, verlinde@phys.uu.nl}
 
 
\end{center}
\vskip3cm

\abstract

We discuss Born-Infeld on the noncommutative two-torus as a description of 
compactified string theory. We show that the resulting  theory, including the 
fluctuations, is manifestly invariant under the T-duality group $\SO(2,2;\Z)$. The 
BPS mass even has a full $\SL(3,\Z)\times\SL(2,\Z)$ U-duality symmetry. 
The direct identification of the noncommutative parameter $\theta$ with the $B$-field 
modulus however seems to be problematic at finite volume. 

\newpage
%
%

\section{Introduction}

String theory compactified on a torus is expected to be invariant under 
a discrete group of U-dualities. An $\SO(n,n;\Z)$ part of this duality group turns 
up as T-duality. This duality is already manifest in the perturbative description 
of string theory. To see the full duality group, also non-perturbative information 
about string theory should be considered. Especially D-branes, which turn up as 
non-perturbative objects in string theory, give necessary states related by U-duality 
to the perturbative string states. In the well-known description of these objects in 
terms of  Dirichlet boundary conditions for (perturbative) strings, the T-duality 
transformations remain valid, and act nontrivially on these objects. 
Already quite some time ago it was found that at low energies the 
world-volume theory of these objects should be described by a Born-Infeld gauge 
theory \cite{lei}. A gauge theory interpretation also turns up when one  looks at 
the M(atrix) theory description of D0-branes \cite{bfss,tay}. There it was found 
that the momentum modes of D-particles describe a gauge theory on the torus dual to 
the one on which these particles live. In this world-volume theory, even the 
T-dualities should act quite nontrivial. For example, for compactification on the 
two-torus there is a duality which interchanges the rank of the gauge group with the 
total magnetic flux. This is a two-dimensional version of Nahm duality.

The dualities act nontrivial on al the moduli of the string theory. One of these 
moduli is the $B$-field. 
Recently it was found that when a nonzero $B$-flux is turned on in string theory, 
the gauge theory describing the D-particles should live on a noncommutative torus. 
this means that the coordinates $x^i$ of the (dual) torus where the gauge theory 
lives should satisfy nontrivial commutation relations (already at the classical level) 
\begin{equation}
[x^i,x^j]=\frac{\theta^{ij}}{2\pi i}.
\end{equation}
It has been argued that the parameter $\theta$ should be identified with a constant 
$B$-field background. This was first discussed in the context of matrix theory in 
\cite{codo}, and discussed thereafter in many papers, see e.g. 
\cite{howu,howu2,dohu,li,cas,sch,kaok,chkr,arar,ho,mozu}. As M(atrix) theory, or more 
correctly DLCQ of M-theory compactified on a torus, 
effectively describes the dynamics of D-branes, this means that also the gauge 
theory on the worldvolume of these branes should be described by a theory on the 
noncommutative torus. 

Combining these known fact, we are led to conjecture that the low energy theory on 
the worldvolume of the D-branes is described by Born-Infeld on a noncommutative torus. 
In this paper we shall consider this proposal for the compactification of string 
theory on a two-torus. This means that we are describing the worldvolume theory on 
the D2-brane. The description using noncommutative geometry has a definite advantage 
above the description using classical geometry. The reason for this is that there is 
a manifest equivalence between gauge bundles over noncommutative tori, called Morita 
equivalence. This equivalence, which actually is an identification of adjoint bundles 
on the torus, has been argued to be the manifestation of T-duality in the gauge theory 
description. In particular, it includes the above mentioned Nahm duality.

In the next subsection, we shortly discuss some well known facts about systems of 
D0-brane on a two-torus and dualities expected from string theory. It will be mainly 
used to set some notations. 

In section \ref{clasBI} we describe the `classical' Born-Infeld theory, which should 
describe these systems, at least for a single D0-brane, in a background with vanishing 
Kalb-Ramond field. We will then discuss a direct generalization of this theory for 
the non-abelian situation with $B\neq0$, in terms of a local field theory. 

In section \ref{NCBI}, a generalization of the Born-Infeld theory living on the 
noncommutative torus will be introduced. The Hamiltonian of this theory will be fixed 
by requiring to have the same dependence on the global zero-modes, as this was already 
correctly described in the naive generalization. We discuss a \emph{manifest} invariance 
of the Hamiltonian under the full T-duality group $\SO(2,2;\Z)$. We also consider the 
Yang-Mills limit of the Born-Infeld theory. 

In section \ref{NCtorus} we give an overview of the noncommutative torus and gauge bundles 
on this geometry. This section will be more technical. The description will be quite 
explicit, in order to make it understandable without much prerequired knowledge. 

Section \ref{concl} concludes with some discussion on what we found.

\subsection{D0-Branes on $\T^2$ and Duality}

In this paper, we are interested in the dynamics of D0-branes in \II{A} string theory 
compactified on a two-dimensional torus $\T^2$. Apart from the D0-branes, we also 
allow for wrapped D2-branes, string winding and momentum. The system can be looked 
at from various points of view. The interpretation of the charges and moduli depend 
on the perspective one takes. From the string point of view, we distinguish the T-dual 
D0- and D2-brane picture. In the two tables \ref{tb:charges} and \ref{tb:moduli} 
we compare the various charges and moduli in these pictures. 

\begin{table}[h]
\begin{minipage}[t]{.56\textwidth}
\begin{center}
\begin{tabular}{|c|c|c|c|}
\hline
charge & D0-brane & D2-brane & Yang-Mills      \\ \hline 
$N$    & D0       & D2       & rank            \\
$M$    & D2       & D0       & $\int\!\Tr F$   \\
$n^i$  & mom.     & F1       & $\int\!\Tr E^i$ \\
$m_i$  & F1       & mom.     & $\int\!\Tr P_i$ \\ \hline
\end{tabular}
\caption{Charges in the various pictures.}\label{tb:charges}
\end{center}
\end{minipage}
\begin{minipage}[t]{.43\textwidth}
\begin{center}
\begin{tabular}{|c|c|c|}
\hline
modulus   & D0-brane    & D2-brane     \\ \hline 
metric    & $g_{ij}$    & $G_{ij}$     \\
$B$-field & $b$         & $B$          \\
coupling  & $\lambda_s$ & $\lambda_s^\prime$ \\
RR-field  & $c_i $      & $C_i$        \\ \hline
\end{tabular}
\caption{Moduli in the various pictures.}\label{tb:moduli}
\end{center}
\end{minipage}
\end{table}

The D2-branes are described by a non-abelian three-dimensional Born-Infeld theory. 
The various quantum numbers in this gauge theory, rank, electric fluxes, magnetic 
flux and momentum, are identified with the charges in the string compactifications, 
as given in table \ref{tb:charges}. 

On the charges and moduli act first of all the T-duality group 
$\SO(2,2;\Z)=\SL(2,\Z)\times\SL(2,\Z)$. Note that the moduli in the D0- and D2-brane 
picture are related by a T-duality transformation. The T-duality group breaks up 
into two $\SL(2,\Z)$ groups. One of them is the mapping class group of the torus, 
which acts trivially on the gauge theory. The other factor acts quite nontrivial. It 
relates the D0-brane picture and the D2-brane picture. This factor will be called 
$\SL(2,\Z)_N$ in the following. One of the generators of this group acts by Nahm 
transformations on the gauge theory. This duality will be our main interest in this 
paper.  

To make the action of T-duality on the moduli more obvious, we express the NS-moduli 
in terms of complex moduli. To be precise, we define a shape parameter $\tau$ and a 
complexified K\"ahler modulus $\sigma$ by 
\begin{equation}
\tau = \frac{g_{12}+i\sqrt{\det g}}{g_{11}}, \qquad 
\sigma = b+i\textstyle\sqrt{\det g},
\end{equation}
The $B$-field modulus $B$ and the volume $\sqrt{\det G}$ in the D2-brane picture are 
related to the latter modulus as $B+i\sqrt{\det G}=-\sigma\inv$. The RR-forms in the 
two pictures are related by Hodge-duality, $c_i=\st C_i=-\eps^{ij}C_j$. The string 
coupling is not invariant under T-duality. The invariant coupling is the effective 
coupling in the eight-dimensional transverse world, given by 
$\lambda_8=\lambda_s/(\det g)^{1/4}=\lambda_s^\prime/(\det G)^{1/4}$.

In the full M-theory, the T-duality group is enhanced to the U-duality group 
$\E_{3(3)}(\Z)=\SL(2,\Z)\times\SL(3,\Z)$. As Type \II{A} string theory can be seen as 
M-theory compactified on a circle, the compactification under consideration is M-theory 
compactified on a three-torus. This explains the second factor of the U-duality group, 
as it is the enhanced mapping class group of the torus. The first factor is hence the 
$\SL(2,\Z)_N$ duality group. The charges in table \ref{tb:charges} transform in the 
$(\mathbf{1},\mathbf{2})\oplus(\mathbf{2},\mathbf{2})$ representation of the T-duality 
group. With the enhancement of the duality group, the charges combine to a 
$(\mathbf{3},\mathbf{2})$ representation of the U-duality group. The moduli (including 
the string coupling) parametrize the moduli space  
\begin{equation}
\CM_3=\frac{\SL(2)}{\SO(2)}\times\frac{\SL(3)}{\SO(3)},
\end{equation}
where the U-duality should also be divided out. This moduli space can be described 
by the $\SL(2)_N$ modulus $\sigma$, and a three-dimensional metric $\widehat G_{\hat\imath\hat\jmath}$, 
build from the remaining moduli $\tau$, $\lambda_8$ and $c_i$. This metric is given by  
\begin{equation}\label{metr3}
\widehat G_{\hat\imath\hat\jmath}=\frac{\ell_{pl}^2}{(\lambda_s^2\det g)^{1/3}}
  \pmatrix{ 
    g_{ij}+\lambda_s^2\,c_ic_j&\lambda_s^2\,c_i\cr 
    \lambda_s^2\,c_j&\lambda_s^2 
  },
\end{equation}
where we introduced the eight-dimensional Planck length $\ell_{pl}=\lambda_8^{1/3}\ell_s$. 
This length unit is invariant under all U-duality transformations, in contrast to the 
string length. Therefore it is appropriate to express all dimensionful quantities that 
are supposed to transform covariantly under the U-duality group in terms of this unit. 
The three-dimensional metric in \eqref{metr3} is not the natural SUGRA metric of M-theory. 
It is rescaled by a factor $\sqrt{\det g}$, which equals the three-dimensional volume of the torus on 
which the M-theory is compactified (in eleven-dimensional Planck units), the only parameter 
available in the M-theory compactification. Note that the factor is such that the metric 
is independent of $\sigma$, so that the moduli space is factorized.

\section{Abelian Born-Infeld on the Classical Torus}
\label{clasBI}

In this section we study the Born-Infeld description on the `classical'
two-torus of Type \II{A} D-branes. At least for the abelian situation arising on single 
two-branes, and with vanishing Kalb-Ramond field, this should be the correct description 
of the effective theory. Starting from this action we shall obtain the full 
BPS spectrum expected from string theory from the Yang-Mills theory, including 
all the moduli.

\subsection{Lagrangian and Hamiltonian}

The correct effective description of D-branes at zero Kalb-Ramond field has been 
known for quite some time \cite{lei}, and is described by a Born-Infeld action.
This action for the abelian gauge theory is given by
\begin{equation}\label{actionbi}
S_{BI} = \int\!d^2xdt\,\frac{-1}{\lambda_s^\prime}\sqrt{\det\bigl(-G-F+B\bigr)} + C\wedge F.
\end{equation}
Here we used units such that the string length is $\ell_s=1$. 
The appearance of the inverse power of the ten-dimensional string coupling 
is natural for the RR-soliton, as a boundary for fundamental strings.
The last term is the coupling to the RR one-form field. There are only fluxes in the two 
spatial torus directions, as these are the only compact directions. Hence this term 
involves only the electric components $F_{0i}$ of the field strength. We included the 
$B$-field as a shift of the field strength, as in \cite{wit}. 

Our next step is to calculate the Hamiltonian associated to the action. This is 
expressed in terms of the field strength, the electric field and the momentum density, 
given by 
\begin{equation}
F=\partial_1\CA_2-\partial_2\CA_1, \qquad
E^i=\frac{\delta\CL}{\delta\dot \CA_i}, \qquad
P_i=F_{ij}E^j.
\end{equation}
The zero-modes of these operators are the various quantum numbers, associated to the 
different branes, windings and momentum, as given in table \ref{tb:charges}. We 
normalize these operators such that the zero-modes are integral. 

Straightforward calculation of the Hamiltonian density then yields the result 
\begin{eqnarray}
\CH &=& E^i\dot \CA_{i}-\CL_{BI} \\
 &=&\frac{1}{\ell_s}\Biggl( \frac{1}{\lambda_8^{2}\sigma_2}|1+\sigma F|^2
 +\frac{1}{\sigma_2\tau_2}\Bigl|-\bar\tau E^1+E^2+\sigma(P_1+\bar\tau P_2)
 +(C_1+\bar\tau C_2)(1+\sigma F)\Bigr|^2 \Biggr)^{1/2},\nonumber
\end{eqnarray}
where in the last line we reintroduced units. 

We now combine the various operators into an `electric' and `magnetic' three-vector. 
This allows us to write the Hamiltonian density in a form where the duality 
transformations are much more obvious, at least as far as the zero-modes are concerned. 
These vectors are 
\begin{equation}\label{EB}
\CE_{\hat\imath}=\pmatrix{ E^i \cr I },\qquad
\CB_{\hat\imath}=\pmatrix{ -\eps^{ij}P_j\cr F }.
\end{equation}
Norms for these vectors should be calculated using the three-dimensional metric \eqref{metr3}. 

Note the position of the indices, which are different because we are comparing 
matrices in T-dual pictures: the objects $\CE$ and $\CB$ should be interpreted in the 
context of the three-dimensional enhancement, which naturally arises starting from the 
D0-brane picture. Using these three-dimensional quantities, including the three-dimensional 
metric \eqref{metr3}, we can write the Hamiltonian density in the suggestive form  
\begin{equation}\label{hamil}
\CH=\frac{\|\CE+\sigma\CB\|}{\ell_{pl}\sqrt{\sigma_2}}.
\end{equation}
Here $\CE$ and $\CB$ have lower indices, hence we used the inverse of the 
metric \eqref{metr3} (with upper indices) to calculate the norm. 

From an M-theory perspective, the above way of writing the Hamiltonian is very natural. 
This is because in M-theory the system is compactified on a three-torus. The vectors 
$\CE$ and $\CB$ are related to the momentum density and membrane-wrapping respectively
($\CB$ is related to a two-vector). In this interpretation, the lower indices are quite 
natural.

\subsection{Naive Generalization and BPS Spectrum}

The result for the Hamiltonian density in the last subsection suggests a 
simple generalization for the non-abelian theory, by replacing the operators 
in $\CE$ and $\CB$ by their non-abelian counterparts and putting a suitable 
trace in front of the expression \eqref{hamil}. It was suggested in \cite{tsey} 
that this trace should be the symmetric trace. Anyway, as far as the zero-modes 
and the BPS masses are concerned, the precise form of the trace is not so relevant.
The zero-modes of the local operators are given by 
\begin{equation}
\CN=\int\!d^2x\Tr\CE=\pmatrix{ n^i\cr N },
\qquad
\CM=\int\!d^2x\Tr\CB=\pmatrix{ -\eps^{ij}m_j\cr M }.
\end{equation}
Inserting these zero-modes in the Hamiltonian \eqref{hamil}, we see that 
this gives the correct zero-mode contribution, invariant under the full 
U-duality group $\SL(2,\Z)\times\SL(3,\Z)$. The BPS mass levels can be calculated 
as a lower bound on the energy. They come out correctly in a manifest U-duality 
invariant way, as they are given by 
\begin{equation}\label{BPSmass}
\ell_{pl}^2M_{BPS}^2= \frac{\|\CN+\sigma\CM\|^2}{\sigma_2} 
 +2\|\CN\times\CM\|.
\end{equation}
This result can be derived methods similar to that used in \cite{hvz} for string 
compactification on the four-torus. In fact, it is a dimensional reduction to 
two dimensions of this result. It equals the direct generalization for $b\neq0$ 
of the result of \cite{dvv} reduced to two dimensions, which was calculated using 
the supersymmetry algebra. Note that this result is certainly correct for $N=1$ 
and $b=0$, as we know that then the Born-Infeld theory is correct. For different 
values of $N$ and $b$, this expression is the unique expression for the BPS mass 
which is invariant under the full expected U-duality. Therefore, whatever theory 
should describe the D0-branes, we know that it must always give rise to these 
BPS masses. In the full supersymmetric Born-Infeld theory, these BPS states are 
states that break half or a quarter of the supersymmetry. For the 1/2 BPS states, 
the mass was already calculated in \cite{eli,piki,blau,obpi} (even for compactification 
on the four-torus). These states correspond to states with  $\CN\times\CM=0$, 
therefore the fluctuations do not contribute to the BPS mass. A nice and recent
review of the subject of U-duality and BPS-formulae in M-theory can be found in 
\cite{obpi2}.

\subsection{Towards Duality Invariance}

Looking at the Hamiltonian density \eqref{hamil}, duality invariance seems quite 
straightforward. Of particular interest in this paper is the $\SL(2,\Z)_N$ duality 
under which the modulus $\sigma$ transforms. Let us look at this duality transformation 
in some more detail. The form of the Hamiltonian \eqref{hamil} suggests that the two 
vectors $\CE$ and $\CB$ transform more or less as a doublet under this $\SL(2,\Z)_N$. 
To guarantee invariance under this group of at least the contribution from the zero-mode 
to the Hamiltonian; the zero-modes should transform exactly as a doublet. This implies 
that the full operators $\CE$ and $\CB$ transform as a doublet up to normalization. The 
normalization can be found from the transformation of the rank $N=\Tr I$.

This simple rule is certainly true at the level of zero-modes. For the complete fields 
including the fluctuations however, this can not be the full story. To see this, 
remark that the unit operator $I$ is interchanged with the magnetic field $F$ under 
the transformation mentioned above. The latter operator however has fluctuations, 
while the unit operator certainly has no fluctuations. Furthermore, the unit operator 
should somehow be the unit operator in all representations. So if we want to view 
duality as some map between different representations, we expect that this operator 
at least should remain intact. And this certainly is not true for this simple 
transformation rule. 

The problem noted above presents itself quite prominent in the metric for the 
fluctuations. To see this, we write down the quadratic action for the fluctuation 
of the magnetic field. For simplicity, and also to make comparison to the 
situation on the noncommutative torus later on, we consider the small volume limit 
$g\to0$. We split the magnetic field into a zero-mode part and a 
fluctuation part, $F=\frac{M}{N}I+F^\prime$. Then by expanding around the zero-modes, 
we find the following quadratic action for the fluctuation part $F^\prime$
\begin{equation}\label{actfluct}
S_{F^\prime}=\frac{N\sigma_2^{3/2}}{\lambda_8\ell_{pl}|N+Mb|}\int\!d^2xdt\,\Tr(F^\prime)^2.
\end{equation}
Let us now see how this transforms under the T-duality transformation 
\begin{equation}
\pmatrix{A&B\cr C&D}\in\SL(2,\Z)_N.
\end{equation}
From the transformation of the zero-modes, we read off that $F^\prime$ has to transform 
according to $NF^\prime\to DNF^\prime$. Furthermore, $\Tr$ transforms as $N=\Tr I$. Also we know 
that the volume $v$ scales in the small volume limit according to 
\begin{equation}
\sigma_2\to\frac{\sigma_2}{(Cb+D)^2}.
\end{equation}
In the small volume limit the zero-mode combination $|N+Mb|/\sqrt{\sigma_2}$ 
is invariant. Combining these transformations, we find that the quadratic 
fluctuation part \eqref{actfluct} transforms with a factor $\frac{D^2}{(Cb+D)^2}$, 
which is different from unity. Therefore, the metric for the fluctuations is not 
invariant. Note that this problem is not a consequence of the generalization to a 
non-abelian gauge theory, because it already happens for transformations that leave 
us in the abelian situation $N=1$. We shall see below how this changes on the 
noncommutative torus.

\section{Born-Infeld on the Noncommutative Torus}
\label{NCBI}

In the last section we found that the non-abelian generalization of the Hamiltonian 
density given in \eqref{hamil} gives rise to the correct BPS masses, and is 
U-duality invariant at the zero-mode level. As we saw however, at the local level 
invariance breaks down when $b$ becomes nonzero. The fact that we successfully reproduced 
the correct BPS masses indicates that the form of the Hamiltonian density can still be 
true, and must be correct for the global modes. We might however try and change the 
identification of the local operators in this formula, to correct for the flaws of the 
theory. This means that we should change the identification of the local forms of 
$\CE$ and $\CB$, but in such a way that they still have the same (integral) zero-modes. 
At first sight this seems  impossible. But here the recent relation of string 
compactifications at nonzero $b$ with noncommutative geometry presents a 
different solution, which, as we shall see in the rest of the paper, will give a solution 
to this problem.

\subsection{Hamiltonian of Born-Infeld on the Noncommutative Torus}

Motivated by the above mentioned observations, we now consider Born-Infeld on the 
noncommutative torus $\T^2_\theta$. As one consequence of this, the fluxes of the 
various operators will change. As noted above, this means that we should replace the 
vectors $\CE$ and $\CB$ of local operators by different ones. These replacements should 
be chosen such that they have the same zero-modes, as we do trust the zero-mode 
contribution to the Hamiltonian. First let us introduce the generalizations of the 
three-vectors of local operators in \eqref{EB} for the noncommutative torus, and denote 
them $\CE_\theta$ and $\CB_\theta$. So e.g. $\CE_\theta=(E^i,I)$. On the noncommutative 
torus, the zero-modes of the various local operators are not integral. They are in 
general shifted, and are given by the formulae \eqref{quantIF} and \eqref{quantEP}; this 
will be discuss this in more detail later. Denoting the zero modes of $\CE_\theta$ and 
$\CB_\theta$ by $\CN_\theta$ and $\CM_\theta$ respectively, these shifts can be 
summarized as 
\begin{equation}
\int\! d^2x\Tr_\theta\CE_\theta\equiv\CN_\theta=\CN+\theta\CM,\qquad 
\int\! d^2x\Tr_\theta\CB_\theta\equiv\CM_\theta=\CM,
\end{equation}
where $\Tr_\theta$ denotes the trace for the bundle on the noncommutative torus. 
These shifts enforce us to identify the three-vectors $\CE$ and $\CB$ appearing in the 
Hamiltonian as shifted fields according to 
\begin{equation}\label{EBth}
\CE=\CE_\theta-\theta\CB_\theta,\qquad \CB=\CB_\theta,
\end{equation} 
in order for them to have integer zero-modes. Inserting these values, we find that the 
Hamiltonian \eqref{hamil} on the noncommutative torus $\T^2_\theta$ can be written 
\begin{equation}\label{hamilnc}
\CH = \frac{\Tr_\theta\|\CE_\theta+(\sigma-\theta)\CB_\theta\|}{\ell_{pl}\sqrt{\sigma_2}}.
\end{equation}
By construction, the zero-mode contribution is exactly the same as for the theory 
on the classical torus. Careful inspection of the derivation shows that even the 
BPS mass spectrum is the same. This also means that this BPS spectrum is invariant 
under the T-duality group. As we shall see below, we have even more.

\subsection{Manifest $\SL(2,\Z)_N$ Duality}

Now that we have constructed the theory on the noncommutative torus, we can start to 
analyze it. We already saw that we have the correct BPS mass levels. Also the $\SL(2,\Z)$ 
part of the T-duality group arising from the mapping class group of the torus  
is manifestly present on the noncommutative torus. The nontrivial part to check is 
the $\SL(2,\Z)_N$ part, which acts on the modulus $\sigma$. 

As we already noted before, the local operators $(\CE,-\CB)$ should transform, modulo 
normalization, as a doublet under this duality group, because the zero-modes of these 
operators do so. In order to find the correct transformation of the zero-modes, the 
local operators should transform according to the rule 
\begin{equation}\label{trafoEB}
\pmatrix{\CE^\prime\cr -\CB^\prime} = 
d \pmatrix{A&B\cr C&D}\pmatrix{\CE\cr -\CB},\qquad
\pmatrix{A&B\cr C&D}\in\SL(2,\Z)_N,
\end{equation}
where $d=\Tr_\theta I/\Tr_{\theta^\prime}I$ takes care of the difference in normalization. 
In noncommutative geometry, it has the interpretation of the scaling of an abstract 
dimension of the fiber of the bundle over the noncommutative torus (which is not 
necessarily integral on the noncommutative torus). To appreciate this, note that for 
the classical torus it equals the scaling of the rank $N$. 

Now let us see what this means for the fields on the noncommutative torus. Before we 
noted that problems arose because the unit operator transforms into a combination of 
the unit operator and the field strength. So let us try to find a transformation such 
that the transformation of the unit operator does not get additional contributions. It 
turns out that, with the transformation rule \eqref{trafoEB} and the identifications 
\eqref{EBth}, the noncommutative parameter $\theta$ must transform with fractional 
linear transformations 
\begin{equation}\label{trafoth}
\theta^\prime = \frac{A\theta+B}{C\theta+D}.
\end{equation}
Using this, we can calculate the scaling of the trace from the formula for the trace of 
the unit
\begin{equation}
d=\frac{N+M\theta}{N^\prime+M^\prime\theta^\prime}=C\theta+D. 
\end{equation}
With the above transformation of $\theta$, the transformation \eqref{trafoEB} of the 
shifted fields can be rewritten in terms of the unshifted variables, related as in 
\eqref{EBth}, as 
\begin{equation}\label{trafoEBth}
\pmatrix{\CE^\prime_{\theta^\prime}\cr -\CB^\prime_{\theta^\prime}} = 
\pmatrix{1&0\cr C(C\theta+D)&(C\theta+D)^2}\pmatrix{\CE_\theta\cr -\CB_\theta}.
\end{equation}
Of particular interest is the zero in the upper right corner of this transformation matrix, 
which assures that the unit operator, which is a component of $\CE_\theta$, does not get 
a contribution from $\CB_\theta$. (The unit operator is even invariant, as the upper left 
element is equal to one. This is a direct consequence of our definition of $d$). The 
above transformation rules can now be consistently written in terms of transformations 
of the independent fields $F$ and $E^i$ according to 
\begin{equation}\label{trafoFE}
F^\prime=(C\theta+D)^2F-C(C\theta+D)I,\qquad 
(E^i)^\prime=E^i.
\end{equation}
As we shall see, the transformation of the field strength $F$ can consistently be written 
in terms of transformation of the gauge field $\CA_i$. These are exactly the transformations 
under Morita-equivalence, which we discuss later. So we find that in our 
Born-Infeld theory on the noncommutative torus, T-duality can be described by Morita 
equivalence.

\subsection{Relation to String Theory}

We shall now discuss the relations of the above description of Born-Infeld theory on the 
noncommutative torus with string compactifications. A direct identification of this theory 
as the effective world-volume theory on the D2-brane runs into an obvious problem: there 
are too many parameters. The string theory has the 7 moduli $g_{ij}$, $b$, $c_i$ and 
$\lambda_s$, while the theory on the noncommutative torus has the additional parameter 
$\theta$. It was argued in several papers, starting from \cite{codo}, that the 
noncommutative parameter $\theta$ should be identified with the $B$-field modulus $b$. 
This has the interesting implication that in the Hamiltonian \eqref{hamilnc} the explicit 
dependence on $b$ drops out; the Hamiltonian only depends on $b$ via the noncommutativity 
encoded in $\theta$. 

This is not the complete solution however. The transformation of $\theta$ and $b$ that we 
derived are different from each other, as $b$ transforms with fractional linear 
transformations together with $\det g$, in the combination $\sigma=b+i\sqrt{\det g}$. 
This has the nasty implication that if we identify $\theta$ with $b$ in some formulation, 
then after doing an $\SL(2,\Z)_N$ duality rotation, the two parameters are not identified 
any more. There can be two ways out of this dilemma. The first is the possible appearance 
of a continuous symmetry which allows us to change $\theta$ to any value; so we can use 
this to just set $\theta$ equal to $b$. This would mean that they are not independent 
parameters, but rather their sum should be identified with the $B$-field. The other way 
out is a deformation of the noncommutative geometry to a torus geometry depending on the 
a complexified modulus $\theta\in\C$. We can then identify $\theta$ with the full complex 
modulus $\sigma$. We believe that the second one is the most plausible. This is because 
we see no reason for the continuous symmetry which otherwise is necessary to exist. Also, 
the generalization of the real parameter $\theta$ in noncommutative geometry to a 
complex one is quite natural, as we know that there is an $\SL(2,\Z)$ duality group 
acting on this parameter; and in most cases where this happens, this parameter can in 
fact be complexified. On the other hand, we do not know such a generalization to exist. 
This solution implies that the description of the noncommutative torus used here as a 
description of string theory compactifications is only correct when $\sigma$ and 
$\theta$ are identified, hence in the small volume limit $g\to0$. If this conjecture is 
true, the identification of $\theta$ with $b$ is only valid in the small volume limit 
$\sigma_2=0$. This is something that, in principle, could be argued from the string 
theory by a refinement of the arguments in the literature that lead to the appearance 
of noncommutative geometry in string compactifications.

\subsection{Lagrangian and Yang-Mills Limit}

We now study the Lagrangian and the Yang-Mills limit of the Born-Infeld theory on the 
noncommutative torus. We will assume here that $\theta$ is identified with the $B$-field 
modulus $b$. The Hamiltonian density on the noncommutative torus that we found can 
easily be derived from a classical Lagrangian. Noting that the form of the Hamiltonian 
density \eqref{hamilnc} is similar to \eqref{hamil} with the explicit $b$-dependence 
taken out, the action from which it should be derived is 
\begin{equation}
S_{BI}=\int\! d^2xdt\,\Tr_\theta\biggl(
  \frac{-1}{g_{YM}^2}\sqrt{\det(-g\inv-F)} + C\wedge F \biggr).
\end{equation}
The coupling that appears in this action should be 
$g_{YM}^2=\lambda_8/(\det g)^{1/4}=\lambda_s/\sqrt{\det g}$, to match the coupling in the Hamiltonian. 
Note that this action in the limit of small coupling, which is the limit where 
noncommutative geometry should appear, reduces to a normal Yang-Mills action. 
The coupling of this Yang-Mills theory is given by $g_{YM}$. 

How can we interpret this result for the action? We see that the metric on the 
noncommutative torus is the inverse metric to the D0-brane metric $g$, and 
not the T-dual metric $G$. This can easily be understood, if we remember that the 
gauge theory arises as the effective theory of the D0-branes represented in terms 
of the momentum modes. Therefore the metric should be the metric on the dual torus 
where the momenta live, and this is the inverse metric. 

Another hint comes from the transformation of the field strength \eqref{trafoFE}. 
The factor $C\theta+D$ is exactly the scaling of the volume $\sqrt{\det g\inv}$ under 
T-duality in the small volume limit $g\to0$, and therefore it is the conformal scaling 
of the metric $g\inv$. This factor can therefore be absorbed by a scaling of the 
coordinates. 

We shall now consider the fluctuation part in the Yang-Mills limit in some more detail. 
The form of the Hamiltonian shows clearly that this limit arises when we take both 
$g$ and $\lambda_8$ to zero. That is, we take the small coupling and small volume 
limit. The Hamiltonian then takes the quadratic form 
\begin{equation}
H = H_0 + \frac{1}{\ell_{pl}\lambda_8\sqrt{\sigma_2}}\int\!d^2xdt\,\Tr_\theta\biggl(
  \frac{\lambda_8^2}{2} (E^\prime)^2+\frac{\sigma_2^2}{2}(F^\prime)^2\biggr),
\end{equation}
where the primes indicate that we omit the zero-modes of the operators.
The first part, $\ell_{pl}H_0=\sigma_2^{-1/2}\|\CN+\sigma\CM\|$, is the zero-mode 
contribution, which is invariant under the complete U-duality group. This 
fluctuation part is now invariant under Morita equivalence or T-duality in the 
small volume limit, as $F^\prime$ scales with $(C\theta+D)^2$, while $\sigma_2$ scales with the 
inverse factor. The overall factor $\sqrt{\sigma_2}\inv$ absorbs the scaling of the 
trace. The coupling $\lambda_8$ and the electric field $E^\prime$ are both invariant. 

As the zero mode part is invariant under the T-duality transformations, it is not 
completely invariant under the small $g$ version, which we saw was equivalent to 
Morita equivalence. Any transformation in this part can however be undone by 
changing the value of certain components of the three-form field, which turn up 
in the Lagrangian by uninteresting constant terms like $\int\Tr_\theta I$ and 
$\int\Tr_\theta F$. This was the point of view taken in \cite{codo} and 
other papers following it. We do however take the point of view that not these 
parameters should be changed, but the transformation rules of the fields 
should be changed to the ones predicted by T-duality. As far as the zero-modes 
are concerned, we already saw that the Hamiltonian is then exactly invariant, at 
least if we take the full Born-Infeld theory. For the fluctuation part, we do not 
know how to correct the transformation rules. Indeed, as Morita equivalence is an 
exact duality for the bundles on the noncommutative torus and not the full duality 
involving $\sigma_2$, the basic definitions of this torus geometry should be changed. 
Somehow, the real parameter $b$ should be generalized to $\sigma$ in the 
complex upper half-plane. 

There is a reason for taking the small volume limit, which is the decoupling of stringy 
corrections\footnote{We thank R. Dijkgraaf for pointing this out to us.}. For this, 
we need to take $\alpha'$ (or $\ell_s$) to zero. Doing this, we should be carefull to 
keep the volume from the gauge theory point of view finite. Otherwise, we can not really 
make sense out of the gauge theory. This volume is the dimensionfull volume 
$\ell_s^2\sqrt{\det g\inv}$, because the natural metric on the noncommutative 
torus is the metric $g\inv$, as we have seen above. Now in the above decoupling limit, 
the finiteness of this volume forces us to take also $g$ to be small. And this is exactly 
the limit in which we can make the identification of the noncommutative parameter $\theta$ 
with the $B$-field modulus $b$. It is also interesting to relate this to the theory on the 
classical torus. The metric there is the T-dual metric $G$. The dimensionfull volume of this 
metric is given by the expression $\ell_s^2\sqrt{\det g}/(b^2+\det g)$. In this paper we are 
interested in a situation where the $B$-field modulus $b$ is finite (of order one). This 
now implies that this T-dual volume always becomes zero in the limit $\alpha'\to0$. Therefore, 
the $\alpha'\to0$ limit, where stringy effects should decouple, would then be described by the 
infrared, free theory. This is however not at all what we should expect. In the description 
using the noncommutative torus however, we see that we can still remain with a theory with 
finite (dimensionfull) in the limit where the stringy effects are decoupled.

\section{Some Notes on the Noncommutative Torus $\T^2_\theta$}
\label{NCtorus}

In this section we briefly discuss the necessary ingredients from noncommutative 
geometry that we need. A quite extensive description of noncommutative geometry in 
general can ber found in Connes' book \cite{con}, where also the noncommutative 
two-torus is discussed in some detail. Some aspects, although in a slightly different 
form, were also discussed in \cite{codo}. The description of the trivial abelian 
bundle was discussed in detail in \cite{howu}.

\subsection{Bundles on the Noncommutative Two-Torus}

The noncommutative two-torus is defined through the $C^*$-algebra of functions, which 
are generated by Fourrier modes $U_i=\e^{2\pi ix^i}$, $i=1,2$. In contrast to the classical 
case, these functions do not commute, but the modes satisfy commutation relations 
\begin{equation}\label{commU}
U_1U_2 = \e^{2\pi i\theta}U_2U_1,\qquad \mbox{or}\qquad 
[x^1,x^2]=\frac{\theta}{2\pi i},
\end{equation} 
where $\theta$ is a real parameter called the noncommutative parameter. Furthermore, 
we need a set of derivations $\partial_i$. They are formally defined through their 
commutation relations with the coordinates, which are the usual ones. Note that they 
are straightforward generalizations of the derivatives, and commute among themselves. 

Gauge bundles on the noncommutative torus are also defined through their sections. As 
these can always be multiplied (on the right) by functions on the torus, they form a 
(right-) \emph{module} for the $C^*$-algebra of functions. This serves in noncommutative 
geometry as the definition of these bundles. Gauge bundles can be endowed with a 
connection $\grad{}$. Connections satisfy the same derivation requirement as in the 
commutative case 
\begin{equation}\label{gradper}
\grad{i}\bigl(\psi f\bigr) = \bigl(\grad{i}\psi\bigr) f + \psi\partial_i f.
\end{equation}
This condition implies that the gauge connection also in the noncommutative case has to  
commute with the full algebra of functions. Therefore, it must be a function not of the 
coordinates $x^i$, but of the modified coordinates $\tilde x^i$ defined as 
\begin{equation}\label{tildeU}
\tilde x^i = x^i+\frac{i\theta}{2\pi}\eps^{ij}\partial_j,
\end{equation}
which commute with the coordinates $x^i$ and therefore with the algebra of functions. 
Note that these form an algebra similar to the $x^i$, but with opposite parameter 
$-\theta$. Gauge bundles can be defined through translation operators defining periodic 
boundary conditions. They are of the form
\begin{equation}
T_i=\e^{\partial_i}\Omega_i\inv. 
\end{equation}
Here the derivative generates the translation, and the $\Omega_i$ are a set of local 
gauge transformations. Gauge transformations should, as they should act only fiberwise, 
be functions of the commuting modes $\tilde x^i$. This also guarantees that the gauge 
field satisfies the derivation condition \eqref{gradper} after a gauge transformation. 

The trivial, abelian, gauge bundle has just the algebra of functions as its sections. The 
periodicity condition are simply determined by $T_i=\e^{\partial_i}$. The gauge 
connection can then be given by $\grad{j}=\partial_j+i\CA_j(\tilde x)$, where the gauge 
field has an expansion in the modes $\widetilde U_i=\e^{2\pi i\tilde x^i}$. Note that 
trivial $\U(N)$ bundles are also easily constructed like this, by simply letting $\CA_j$ 
take values in the corresponding Lie-algebra. 

We come now to the construction of non-abelian gauge bundles with gauge group $\U(N)$ 
and with nonzero magnetic flux $M$. We assume that $N$ and $M$ are relatively prime 
numbers. We can always take a tensor product with a trivial $\U(N_0)$-bundle to describe 
the general case. A $\U(N)$-bundle with flux $M$ can be constructed following 't Hoofts 
approach, starting from $\SU(N)$ matrices 
satisfying 
\begin{equation}
V_1V_2=\e^{2\pi i\frac{M}{N}}V_2V_1.
\end{equation}
They can be constructed as suitable `clock and shift' matrices. 
The translation operators defining the periodicity conditions can then be given by 
\begin{equation}
T_1=\e^{\partial_1}V_1,\qquad 
T_2=\e^{\partial_2}\e^{-\frac{2\pi iM}{N}\tilde x^1}V_2,
\end{equation}
Here an abelian gauge transformation is added to assure that they commute.\footnote{This is 
necessary in order for a fundamental bundle to exist. In the string interpretation, we
need fundamentals because string endpoints, which should be allowed for D-branes, are 
seen as fundamentals on the worldvolume of the D-brane.}

The gauge field should always commute with both the coordinates $x^i$ and the
translations $T_i$. This turns out to imply that they can be expanded in the modes 
\begin{equation}\label{Zdef}
Z_1 = \e^{\frac{2\pi i}{N}\tilde x^1}V_2^{K},\qquad
Z_2 = \e^{\frac{2\pi i}{N+M\theta}\tilde x^2}V_1^{-K},
\end{equation}
where $K$ is an integer satisfying $NL-KM=1$ for some integer $L$. Note that the 
relative prime condition implies the existence of such an integer. These modes are 
the non-abelian generalizations of the modes $\widetilde U_i$. They generate an algebra 
similar to that of the $\widetilde U_i$ 
\begin{equation}\label{commZ}
Z_1Z_2=\e^{-2\pi i\hat\theta}Z_2Z_1,
\qquad \mbox{where}\qquad
\hat\theta = \frac{K+L\theta}{N+M\theta}.
\end{equation}
We can construct a linear covariant derivative by the ansatz 
\begin{equation}
\exp{\grad[0]{1}}=T_1Z_2^{-M},\qquad 
\exp{\grad[0]{2}}=T_2Z_1^{M}.
\end{equation}
Here the exponent $M$ is chosen such that the $\SU(N)$ matrices drop out, so that we can 
take the logarithm. The presence of the $T_i$ guarantee the correct derivation property 
\eqref{gradper}. The general connection can then written $\grad{j}=\grad[0]{j}+i\CA_j(Z)$. 
From the linear connection above we can calculate the zero-mode of the field strength 
\begin{equation} 
F^0=\frac{1}{2\pi i}[\grad[0]{1},\grad[0]{2}]=\frac{M}{N+M\theta}I.
\end{equation} 
Note the different normalization of this field strength. This is a consequence of the 
fact that the canonical trace on this bundle is not normalized to $N$, but rather \cite{con}
\begin{equation}\label{quantIF}
\Tr_\theta I=N+M\theta,\qquad\mbox{with}\qquad
\int\!d^2x\Tr_\theta F=M.
\end{equation}
Note that the magnetic flux is still an integer.

\subsection{$\SL(2,\Z)_N$ Duality: Morita Equivalence}

we now give a short discussion of Morita equivalence, which is a mathematical equivalence 
for bundles on the noncommutative torus. In the application to string theory, this duality 
is directly related to T-duality. 

We saw above that the modes $Z_i$ of the non-abelian gauge field generate an algebra 
\eqref{commZ} of the same type as the abelian modes $\widetilde U_i$. Therefore, we can 
identify the modes $Z_i$ with modes of an abelian gauge field. This identifies the 
non-abelian gauge field $\CA$ with an abelian gauge field $\widehat\CA$. The corresponding 
abelian gauge bundle lives on a different noncommutative torus. Comparing the commutation 
relations of the $Z_i$ with those of the $\widetilde U_i$, we see that this should be a 
torus with noncommutative parameter $\hat\theta$. So we find that the \emph{non-abelian} 
gauge field $\CA$ on the non-commutative torus $\T^2_\theta$ can be identified with a 
\emph{trivial abelian} gauge field $\widehat\CA$ on the dual torus $\widehat\T^2_{\hat\theta}$.
Note that two gauge theories are in fact related by an $\SL(2,\Z)$ transformation, with 
matrix
\begin{equation}\label{klmnmat}
\pmatrix{L&K\cr M&N}\in\SL(2,\Z).
\end{equation}
Indeed, the transformed noncommutative parameter $\hat\theta$ in \eqref{commZ} is related by 
the above matrix through standard fractional linear transformations. The pair $(N,-M)$ 
is mapped by this matrix to $(1,0)$, which is the gauge data appropriate for the trivial 
abelian bundle. The only thing still missing to get a direct mapping between the gauge fields 
is the normalization of the fields. 
To find the correct normalization we look at the linear connection $\grad[0]{}$. 
The dual linear connection is the trivial one $\widehat\partial$. Taking care of the 
scaling of the coordinates, which can be read off from the explicit form of the $Z_i$, 
we find the explicit relation between the linear connections 
\begin{equation}
\hatgrad[0]{1} = \widehat\partial_1 = (N+M\theta)\grad[0]{1}+2\pi iMx^2,\qquad 
\hatgrad[0]{2} = \widehat\partial_2 = (N+M\theta)\grad[0]{2}.
\end{equation}
This implies a scaling of the gauge field with a factor of $(N+M\theta)$. The abelian 
field strength is then easily found in terms of the non-abelian one 
\begin{equation}
\widehat F = (N+M\theta)^2F -M(N+M\theta)I.
\end{equation}
Note that the linear shift absorbs the zero-mode $F^0$ of the non-abelian field 
strength, so that indeed the abelian gauge bundle is trivial. 

We saw that we can always relate a non-abelian gauge field on the noncommutative torus 
with an abelian gauge field on a dual torus. We can turn this around, and relate the 
abelian gauge field to several non-abelian gauge fields on different tori. This 
correspondence defines an equivalence called \emph{Morita equivalence}. It is an 
equivalence relation between bundles on noncommutative tori. We saw that in fact it is 
an isomorphism of the corresponding adjoint bundles, which are generated by the $Z_i$. 
The above equivalence between the non-abelian bundle and the abelian one can be 
generated by a unimodular matrix \eqref{klmnmat}.
It was already noted that this matrix acts on the parameter $\theta$ to find $\hat\theta$. 
Moreover, it transforms the vector of charges $(N,-M)$, which characterize the non-abelian 
bundle, into the vector $(1,0)$, which then stands for the trivial abelian bundle. 
Combining several of these equivalences, this implies that any of these can be generated 
by an $\SL(2,\Z)$ transformation, acting on the bundle data $(N,-M)$ in the doublet 
representation, and on the noncommutative parameter $\theta$ with fractional linear 
transformations. Explicitly, these transformations act as follows 
\begin{eqnarray}\label{trafonc}
&& \CA\to(C\theta+D)\CA,\qquad 
F\to(C\theta+D)^2F-C(C\theta+D),\nonumber\\
&& \theta\to\frac{A\theta+B}{C\theta+D},\qquad
\mbox{for}\quad \pmatrix{A&B\cr C&D}\in\SL(2,\Z).
\end{eqnarray}

We already noted that the canonical trace in a bundle on the noncommutative torus is 
given by the non-trivial expression $N+M\theta$. Also remember that the adjoint bundles 
in the dual situations can be identified. Therefore the trace in one representation also 
defines a trace in the dual one. Of course, a trace can always be multiplied by a 
constant. As the trace has a canonical normalization, where the magnetic field has an 
integer zero-mode, the factor with which the trace changes can be found. The easiest 
way is by considering the formula for the trace of the unit in \eqref{quantIF}. With the 
action of $\SL(2,\Z)$ on $\theta$ and on the doublet $(N,-M)$, we find 
\begin{equation}
\Tr_\theta\to(C\theta+D)\inv\Tr_\theta.
\end{equation}

To finish the discussion about the duality, we consider the transformation of the 
electric field $E^i=\delta\CL/\delta\dot\CA_i$. The transformation can now most directly 
be read off from the invariance of the combination $\Tr_\theta E^i\dot \CA_i$. It follows 
that $E^i$ is invariant.

\subsection{Quantization of the Electric Flux}

We now study the periodicity of the gauge field. As the $\U(N)$ gauge field takes values in 
a compact space, there are certain periodicity conditions. Most important is the $\U(1)$ 
factor, which gives a periodicity $\CA_i\to \CA_i+2\pi$. This periodicity is generated 
by the electric field operator $E^i=\delta/\delta \CA_i$. On the commutative torus this 
shift is generated by a gauge transformation, with $\Omega=\exp 2\pi ix^i$. On the 
noncommutative torus this is not a local gauge transition function, as we saw above. The 
gauge transformation that is most close to this on the noncommutative torus is 
$\Omega=\exp(2\pi ix^i+\theta^{ij}\grad{j})$. Note that this is a function of the $\tilde x^i$, 
if we combine the derivative in $\grad{}$ with the coordinates. Also note that this 
gauge transformation commutes with the $T_i$. Therefore, it is a global section of the 
principal fibre bundle related to the gauge bundle, which is a necessary condition for 
global gauge transformations. We find 
\begin{equation}
\grad{j}\to\e^{\theta^{ik}\partial_k}\grad{j}\e^{-\theta^{ik}\grad{k}}-2\pi i\delta_{ij}.
\end{equation}
Now the shift part is generated by the electric field zero mode $\int\!\Tr E^i$, while 
the covariant derivative generates a translation, which can be interpreted as the action of  
the total momentum operator, $\int\!\Tr_\theta P_i$. The gauge transformation on the gauge 
field above is therefore generated on the wave function by the operator 
\begin{equation}
\exp\Bigl(2\pi i\int\Tr_\theta(E^i+\theta^{ik}P_k)\Bigr).
\end{equation}
As it is a true gauge transformation ($\Omega$ is single valued on the noncommutative 
torus), this operation should act trivial on the wave function. Therefore, the 
quantization of the electric flux is modified on the noncommutative torus to 
\begin{equation}\label{quantEP}
\int\!\Tr_\theta E^i = n^i-\theta^{ij}m_j,\qquad \mbox{where}\qquad
\int\!\Tr_\theta P_i = m_i.
\end{equation}
Here both $n^i$ and $m_i$ are integers. 
Note that the total momentum is still quantized in the usual way, because they are related 
to the periodicity of the torus.

\section{Discussion and Conclusion}
\label{concl}

We found that the Born-Infeld theory on the dual noncommutative torus gives the 
correct U-duality invariant BPS spectrum expected from string theory and M-theory. 

This was also true for the theory on the commutative torus, when the constant 
$B$-field was included in the action in a different way. But when put on a 
noncommutative torus, it turn out that the map corresponding to the T-duality part 
of the duality group can be extended to an exact map between the different gauge 
theories. This map turns out to be given by Morita equivalence. This was already 
appreciated in \cite{codo}, and generalized in \cite{sch} for the M(atrix) theory. 
We found however that this is only consistent with the identification of the 
$B$-field modulus when the volume $\sigma_2$ is zero. We argued that for finite 
values of this volume, there should be corrections to the noncommutative geometry, 
which enable us to identify the noncommutative parameter $\theta$ with the full 
complex modulus $\sigma$. However, as already mentioned, such a generalization is 
not known to us at present; although it is very natural that it should exist. 
On the other hand, we argued that the fact that the dimensionless volume is zero 
is related to the supression of stringy effects. So we expect that for a description 
at nonzero $\sigma_2$ we have to include also the stringy effects, which might modify 
the theory considerably. The gauge theory description then may not be sufficient. 

The nontrivial $\SL(2,\Z)_N$ duality group is generated by two operations. One is a 
shift transformation, which maps the magnetic flux $M$ to $M+N$. This duality is quite 
trivial in the gauge theory. The other one, which 
interchanges the magnetic flux $M$ and the rank $N$, is much more nontrivial, and is 
a two-dimensional version of the Nahm transformation, described in \cite{have}. In 
the `classical' description on the commutative torus, this duality is quite involved; 
it is formulated in terms of zero-modes of the Dirac operator. In the setting of 
noncommutative geometry however, this transformation is much simpler. As we saw, 
the adjoint fields on both sides can really be identified. This identification can
even be done locally, as far as this can be said in noncommutative geometry.

The U-duality transformations that are not part of the T-duality group are still a bit 
mysterious. They are only seen at the level of the BPS spectrum in the gauge theory. 
In \cite{have}, it was noted that for Yang-Mills on the three-torus a subset of the 
BPS spectrum even has the correct U-duality invariant counting formula. This result 
can directly be reduced to two dimensions. Because we identified the Yang-Mills theory 
as a limit of the Born-Infeld theory, we expect also the counting of the corresponding 
BPS states in the Born-Infeld theory to be U-duality invariant.

\ack

It is a pleasure to thank Gijsbert Zwart, Feike Hacquebord, Robbert Dikgraaf 
and Adriene Criscuolo for interesting discussions and helpful comments. C.H. is 
financially supported by the Stichting FOM. The research of E.V. is partly 
supported by the Pionier Programme of the Netherlands Organisation for Scientific 
Research (NWO).

\end{document}